\documentclass[aps,prd,twocolumn,superscriptaddress,showpacs]{revtex4}
\usepackage[dvips]{graphicx}
\usepackage{amsmath,amssymb,times}

\newcommand{\bequ}{\begin{equation}}
\newcommand{\eequ}{\end{equation}}
\newcommand{\bea}{\begin{eqnarray}}
\newcommand{\eea}{\end{eqnarray}}

\DeclareSymbolFont{boldletters}{OML}{cmm} {b}{it}
\DeclareSymbolFontAlphabet{\mathbit}{boldletters}
\DeclareMathSymbol{\alpha}{\mathalpha}{letters}{"0B}
\DeclareMathSymbol{\beta}{\mathalpha}{letters}{"0C}
\DeclareMathSymbol{\gamma}{\mathalpha}{letters}{"0D}
\DeclareMathSymbol{\delta}{\mathalpha}{letters}{"0E}
\DeclareMathSymbol{\epsilon}{\mathalpha}{letters}{"0F}
\DeclareMathSymbol{\zeta}{\mathalpha}{letters}{"10}
\DeclareMathSymbol{\eta}{\mathalpha}{letters}{"11}
\DeclareMathSymbol{\theta}{\mathalpha}{letters}{"12}
\DeclareMathSymbol{\iota}{\mathalpha}{letters}{"13}
\DeclareMathSymbol{\kappa}{\mathalpha}{letters}{"14}
\DeclareMathSymbol{\lambda}{\mathalpha}{letters}{"15}
\DeclareMathSymbol{\mu}{\mathalpha}{letters}{"16}
\DeclareMathSymbol{\nu}{\mathalpha}{letters}{"17}
\DeclareMathSymbol{\xi}{\mathalpha}{letters}{"18}
\DeclareMathSymbol{\pi}{\mathalpha}{letters}{"19}
\DeclareMathSymbol{\rho}{\mathalpha}{letters}{"1A}
\DeclareMathSymbol{\sigma}{\mathalpha}{letters}{"1B}
\DeclareMathSymbol{\tau}{\mathalpha}{letters}{"1C}
\DeclareMathSymbol{\upsilon}{\mathalpha}{letters}{"1D}
\DeclareMathSymbol{\phi}{\mathalpha}{letters}{"1E}
\DeclareMathSymbol{\chi}{\mathalpha}{letters}{"1F}
\DeclareMathSymbol{\psi}{\mathalpha}{letters}{"20}
\DeclareMathSymbol{\omega}{\mathalpha}{letters}{"21}
\DeclareMathSymbol{\varepsilon}{\mathalpha}{letters}{"22}
\DeclareMathSymbol{\vartheta}{\mathalpha}{letters}{"23}
\DeclareMathSymbol{\varpi}{\mathalpha}{letters}{"24}
\DeclareMathSymbol{\varrho}{\mathalpha}{letters}{"25}
\DeclareMathSymbol{\varsigma}{\mathalpha}{letters}{"26}
\DeclareMathSymbol{\varphi}{\mathalpha}{letters}{"27}
\DeclareMathSymbol{\Gamma}{\mathalpha}{letters}{"00}
\DeclareMathSymbol{\Delta}{\mathalpha}{letters}{"01}
\DeclareMathSymbol{\Theta}{\mathalpha}{letters}{"02}
\DeclareMathSymbol{\Lambda}{\mathalpha}{letters}{"03}
\DeclareMathSymbol{\Xi}{\mathalpha}{letters}{"04}
\DeclareMathSymbol{\Pi}{\mathalpha}{letters}{"05}
\DeclareMathSymbol{\Sigma}{\mathalpha}{letters}{"06}
\DeclareMathSymbol{\Upsilon}{\mathalpha}{letters}{"07}
\DeclareMathSymbol{\Phi}{\mathalpha}{letters}{"08}
\DeclareMathSymbol{\Psi}{\mathalpha}{letters}{"09}
\DeclareMathSymbol{\Omega}{\mathalpha}{letters}{"0A}

\begin{document}
\title{Theta vacuum and entanglement interaction \\
in the three-flavor Polyakov-loop extended Nambu-Jona-Lasinio model}

\author{Takahiro Sasaki}
\email[]{sasaki@phys.kyushu-u.ac.jp}
\affiliation{Department of Physics, Graduate School of Sciences, Kyushu University,
             Fukuoka 812-8581, Japan}

\author{Junichi Takahashi}
\email[]{takahashi@phys.kyushu-u.ac.jp}
\affiliation{Department of Physics, Graduate School of Sciences, Kyushu University,
             Fukuoka 812-8581, Japan}

\author{Yuji Sakai}
\email[]{sakai@phys.kyushu-u.ac.jp}
\affiliation{Department of Physics, Graduate School of Sciences, Kyushu University,
             Fukuoka 812-8581, Japan}

\author{Hiroaki Kouno}
\email[]{kounoh@cc.saga-u.ac.jp}
\affiliation{Department of Physics, Saga University,
             Saga 840-8502, Japan}

\author{Masanobu Yahiro}
\email[]{yahiro@phys.kyushu-u.ac.jp}
\affiliation{Department of Physics, Graduate School of Sciences, Kyushu University,
             Fukuoka 812-8581, Japan}

\date{\today}

\begin{abstract}
We investigate theta-vacuum effects on the QCD phase diagram for 
the realistic 2+1 flavor system, using 
the three-flavor Polyakov-extended Nambu-Jona-Lasinio (PNJL) model 
and the entanglement PNJL model as an extension of the PNJL model. 
The theta-vacuum effects make the chiral transition sharper. 
For large theta-vacuum angle the chiral transition becomes first order 
even if the quark number chemical potential is zero, 
when the entanglement coupling between the chiral condensate 
and the Polyakov loop is taken into account. 
We finally propose a way of circumventing the sign problem on lattice QCD with 
finite theta. 
\end{abstract}

\maketitle
\section{Introduction} 

The existence of instanton solution in quantum chromodynamics (QCD) 
requires the QCD Lagrangian with the theta vacuum: 
\bea
{\cal L}_{QCD}
&=&
\bar{q}_f(\gamma_{\nu}D_{\nu}+m_f)q_f
+\frac{1}{4}F^{a}_{\mu\nu}F^{a}_{\mu\nu} 
\nonumber \\
&~&-i\theta\frac{g^2}{64\pi^2}\epsilon_{\mu\nu\sigma\rho}F^{a}_{\mu\nu}F^{a}_{\sigma\rho} 
\eea
in Euclidean spacetime, where 
$F^{a}_{\mu\nu}$ is the field strength of gluon~\cite{Theta76}. 
Theoretically the angle 
$\theta$ can take any arbitrary value 
between $-\pi$ and $\pi$.  
However, experiments indicate 
$|\theta| < 3 \times 10^{-10}$~\cite{Baker,Ohta}. 
The Lagrangian is invariant under the combination of 
the parity transformation and the transformation $\theta \to -\theta$, 
so that parity ($P$) and charge-parity symmetry ($CP$) become 
exact only at $\theta=0$ and $\pm \pi$; note that 
$\theta=-\pi$ is identical with $\theta=\pi$ because of the periodicity 
$2\pi$ in $\theta$. 
Why should $\theta$ be so small? This long-standing puzzle is called 
the strong $CP$ problem; 
see Ref.~\cite{Vicari} and references therein for the detail. 

For $T$ higher than the QCD scale $\Lambda_{\mathrm QCD}$, 
there is a possibility that $\theta$ is effectively varied 
to finite values depending on spacetime coordinates $(t,x)$, 
since sphalerons are so activated as to jump over the potential 
barrier between the different degenerate ground states~\cite{MMS}. 
If this happens, $P$ and $CP$ symmetries can be violated locally 
in high-energy heavy-ion collisions or the early Universe 
at $T \approx \Lambda_{\mathrm QCD}$. 
Actually, it is argued in Refs.~\cite{MZ,KZ} that $\theta$ may be 
of order 1 at the epoch of the QCD phase transition in the early Universe, 
whereas it vanishes at the present 
epoch~\cite{Peccei,Dine,Zhitnitsky,Shifman,Kim}. 
This finite value of $\theta$ could be a new source of large $CP$ violation 
in the early Universe and may be a crucial missing element 
for solving the puzzle of baryogenesis. 

In the early stage of heavy-ion collision, the magnetic field is formed, while 
the effective $\theta (t,x)$ deviates 
the total number of particles plus antiparticles with right-handed 
helicity from those with left-handed helicity. 
As a consequence of this fact, 
an electromagnetic current is generated along the magnetic field, 
since particles with right-handed helicity move opposite 
to antiparticles with right-handed helicity. 
This is the so-called chiral magnetic effect~\cite{Kharzeev,KZ,FKW,Fukushima3}. The chiral magnetic effect may explain the charge separations 
observed in the recent STAR results~\cite{Abelev}. 
The thermal system with nonzero $\theta$ is thus quite interesting. 

For vacuum with no temperature ($T$) and 
no quark-number chemical potential ($\mu$), 
parity $P$ is preserved when $\theta =0$~\cite{VW}, but is 
spontaneously broken when $\theta =\pi$~\cite{Dashen,Witten}. 
The $P$ violation, called the Dashen mechanism, 
is essentially nonperturbative, but the first-principle lattice QCD (LQCD) 
is not applicable for the case of finite $\theta$ because of the sign problem.
Temperature ($T$) and/or quark-number chemical potential ($\mu$) 
dependence of the mechanism has then been analyzed by effective models 
such as the chiral perturbation theory~\cite{VV,Smilga,Tytgat,ALS,Creutz,MZ}, 
the Nambu-Jona-Lasinio (NJL) model~\cite{FIK,Boer,Boomsma,Chatteriee}  
and the Polyakov-loop extended Nambu-Jona-Lasinio (PNJL) 
model~\cite{Kouno_CP,Sakai_CP}. 

Using the two-flavor NJL model~\cite{NJ1,AY,Kashiwa}, 
Fujihara, Inagaki and Kimura made a pioneering work 
on the $P$ violation at $\theta=\pi$~\cite{FIK} 
and Boer and Boomsma studied this issue extensively~\cite{Boer,Boomsma}. 
In the previous works~\cite{Kouno_CP,Sakai_CP}, we extended the formalism 
to the two-flavor PNJL and entanglement PNJL (EPNJL) models and 
investigated effects of the theta vacuum on 
the QCD phase diagram.  Very recently similar analyses were made 
for the realistic case of 2+1 flavors 
by using the NJL model~\cite{Chatteriee}. 
It is then highly expected that the finite-$\theta$ effect is 
investigated in the 2+1 flavor case by using the PNJL and EPNJL models 
that are more reliable than the NJL model. 

If QCD with finite $\theta$ is analyzed directly with LQCD, one can get 
conclusive results on the theta vacuum. 
Here we consider a way of minimizing the sign problem in LQCD. 
For this purpose we transform the quark field $q$ to 
the new one $q^\prime$ 
by the following $SU_{\rm A}(3) \otimes U_{\rm A}(1)$ transformation, 
\begin{equation}
q_u=e^{i\gamma_5{\theta\over{4}}}q_u^\prime ,~~~~~
q_d=e^{i\gamma_5{\theta\over{4}}}q_d^\prime ,~~~~~
q_s=q_s^\prime .
\label{UAtrans}
\end{equation}
The QCD Lagrangian is then rewritten into 
\begin{equation}
{\cal L}_{QCD}
=
\sum_{l=u,d}
\bar{q}_l'{\cal M}_l(\theta )q_l'
+\bar{q}_s'{\cal M}_sq_s'
+\frac{1}{4}F^{a}_{\mu\nu}F^{a}_{\mu\nu}
\label{QCD-2}
\end{equation}
with the new quark field $q^\prime$, where 
\begin{eqnarray}
{\cal M}_l(\theta )&\equiv&\gamma_{\nu}D_{\nu}+m_l\cos{(\theta /2)}+m_li\gamma_5\sin{(\theta /2)},
\\
{\cal M}_s&\equiv&\gamma_{\nu}D_{\nu}+m_s.
\end{eqnarray}
Only the Dirac operator ${\cal M}_l(\theta)$ has $\theta$ dependence 
in \eqref{QCD-2}. 
The determinant of ${\cal M}_l(\theta)$ satisfies 
\begin{eqnarray}
{\rm det}{\cal M}_l(\theta)
=
\left(
{\rm det}{\cal M}_l(-\theta)
\right)^{\ast}.
\end{eqnarray}
The sign problem is thus induced by the $\theta$-odd ($P$-odd) term, 
$m_l i\gamma_5 \sin{(\theta /2)}$. 
The difficulty of the sign problem is expected to be minimized in 
the QCD Lagrangian of \eqref{QCD-2}, since the $\theta$-odd term includes 
the light quark mass $m_l$ that is much smaller than $\Lambda_{\rm QCD}$ 
as a typical scale of QCD. This point is discussed in this paper. 

In this paper, we analyze effects of the theta vacuum 
on the QCD phase diagram for the realistic case of 2+1 flavors by using 
the three-flavor PNJL~\cite{Matsumoto} and EPNJL models~\cite{Sasaki-T_Nf3}. 
Particularly, the three-flavor phase diagram is investigated 
in the $\mu$-$T$ plane with some values of $\theta$. 
Through the analysis, we finally propose a way of circumventing the 
sign problem on LQCD calculations with finite $\theta$. 

This paper is organized as follows. 
In Sec.~\ref{Formalism}, we recapitulate the three-flavor PNJL and EPNJL 
models. 
In Sec.~\ref{Numerical results}, numerical results are shown.  
Section \ref{Summary} is devoted to summary.   

\section{Formalism}
\label{Formalism}
The three-flavor PNJL Lagrangian with the $\theta$-dependent anomaly term 
is obtained in Euclidean spacetime by 
\begin{align}
 {\cal L}  
=& {\bar q}(\gamma_\nu D_\nu + {\hat m_0} - \gamma_4 {\hat \mu} )q  
\nonumber\\
 &- G_{\rm S} \sum_{a=0}^{8} 
    [({\bar q} \lambda_a q )^2 +({\bar q }i\gamma_5 \lambda_a q )^2] 
\nonumber\\
 &+ G_{\rm D} \Bigl[e^{i\theta}\det_{ij} {\bar q}_i (1-\gamma_5) q_j 
           +e^{-i\theta}\det_{ij} {\bar q}_i (1+\gamma_5) q_j \Bigr]
\nonumber\\
&+{\cal U}(\Phi [A],\Phi^* [A],T) , 
\label{L_vacuum}
\end{align} 
where $D_\nu=\partial_\nu - i\delta_{\nu 4}A_{4}^{a}{\lambda_a /2}$ 
with the Gell-Mann matrices $\lambda_a$. 
The corresponding PNJL Lagrangian in Minkowski spacetime is shown 
in Refs.~\cite{Kouno_CP,Sakai_CP}. 
The three-flavor quark fields $q=(q_u,q_d,q_s)$ have masses 
${\hat m_0}={\rm diag}(m_u,m_d,m_s)$, and the chemical potential 
matrix ${\hat \mu}$ is defined by ${\hat \mu}={\rm diag}(\mu,\mu,\mu)$ 
with the quark-number chemical potential $\mu$. 
Parameters $G_{\rm S}$ and $G_{\rm D}$ denote coupling constants 
of the scalar-type four-quark and the Kobayashi-Maskawa-'t Hooft (KMT) determinant interaction~\cite{KMK,tHooft}, respectively, where 
the determinant runs in the flavor space. 
The KMT determinant interaction breaks 
the $U_\mathrm{A} (1)$ symmetry explicitly. 
Obviously, the theta-vacuum parameter $\theta$ has a periodicity of $2\pi$. 
We then restrict $\theta$ in a period $0 \le \theta \le 2\pi$. 

The gauge field $A_\mu$ is treated as a homogeneous and static 
background field in the PNJL model~\cite{Meisinger,Fukushima,Ratti,Rossner,Schaefer,Kashiwa1,Sakai,Sakai2, Kahara,Kashiwa5,Kouno,Matsumoto,Sasaki-T,Sakai_phase,Sakai5,Gatto,Kouno_CP,Sasaki-T_Nf3,Sakai_CP}. 
The Polyakov-loop $\Phi$ and its conjugate $\Phi ^*$ are determined 
in the Euclidean space by
\begin{align}
\Phi &= {1\over{3}}{\rm tr}_{\rm c}(L),
~~~~~\Phi^* ={1\over{3}}{\rm tr}_{\rm c}({\bar L}),
\label{Polyakov}
\end{align}
where $L  = \exp(i A_4/T)$ with $A_4/T={\rm diag}(\phi_r,\phi_g,\phi_b)$ 
in the Polyakov-gauge; note that 
$\lambda_a$ is traceless and hence $\phi_r+\phi_g+\phi_b=0$. 
Therefore we obtain 
\begin{eqnarray}
\Phi &=&{1\over{3}}(e^{i\phi_r}+e^{i\phi_g}+e^{i\phi_b})
\notag\\
&=&{1\over{3}}(e^{i\phi_r}+e^{i\phi_g}+e^{-i(\phi_r+\phi_g)}), 
\notag\\
\Phi^* &=&{1\over{3}}(e^{-i\phi_r}+e^{-i\phi_g}+e^{-i\phi_b})
\notag\\
&=&{1\over{3}}(e^{-i\phi_r}+e^{-i\phi_g}+e^{i(\phi_r+\phi_g)}) .
\label{Polyakov_explict}
\end{eqnarray}
We use the Polyakov potential $\mathcal{U}$ of Ref.~\cite{Rossner}: 
\begin{align}
&{\cal U} = T^4 \Bigl[-\frac{a(T)}{2} {\Phi}^*\Phi\notag\\
      &~~~~~+ b(T)\ln(1 - 6{\Phi\Phi^*}  + 4(\Phi^3+{\Phi^*}^3)
            - 3(\Phi\Phi^*)^2 )\Bigr]
            \label{eq:E13}
\end{align}
with 
\begin{align}
a(T)   = a_0 + a_1\Bigl(\frac{T_0}{T}\Bigr)
                 + a_2\Bigl(\frac{T_0}{T}\Bigr)^2,~~~~
b(T)=b_3\Bigl(\frac{T_0}{T}\Bigr)^3 .
            \label{eq:E14}
\end{align}
The parameter set in $\mathcal{U}$ is fitted to LQCD data at finite $T$ 
in the pure gauge limit. 
The parameters except $T_0$ are summarized in Table \ref{table-para}.  
The Polyakov potential yields a first-order deconfinement phase transition 
at $T=T_0$ in the pure gauge theory. The original value of $T_0$ is 
$270$~MeV determined from the pure gauge LQCD data, 
but the PNJL model with this value of $T_0$ yields a larger 
value of the pseudocritical temperature $T_\mathrm{c}$ of the deconfinement transition 
at zero chemical potential than $T_{\rm c}\approx 160$~MeV predicted 
by full LQCD \cite{Borsanyi,Soeldner,Kanaya}. 
Therefore we rescale $T_0$ to 195~(150)~MeV 
so that the PNJL (EPNJL) model can reproduce $T_{\rm c}=160$~MeV~\cite{Sasaki-T_Nf3}. 
\begin{table}[h]
\begin{center}
\begin{tabular}{cccc}
\hline \hline
$a_0$ & $a_1$ & $a_2$ & $b_3$
\\
\hline
~~~$3.51$~~ & ~~$-2.47$~~ & ~~$15.2$~~~ & ~~$-1.75$~~\\
\hline \hline
\end{tabular}
\caption{
Summary of the parameter set in the Polyakov-potential sector 
determined in Ref.~\cite{Rossner}. 
All parameters are dimensionless. 
}
\label{table-para}
\end{center}
\end{table}

Now the quark field $q$ is transformed into the new one $q^\prime$ 
by \eqref{UAtrans} 
in order to remove $\theta$ dependence of the determinant interaction. 
As shown later, this transformation provides 
the thermodynamic potential $\Omega$ with 
a compact form and furthermore convenient to discuss the sign problem in 
LQCD. 
The present three-flavor PNJL model has 18 scalar and 
pseudoscalar condensates of quark-antiquark pair, but 
flavor off-diagonal condensates vanish for the system with 
flavor symmetric chemical potentials 
only~\cite{Boer,Boomsma,Kouno_CP,Sakai_CP}. 
Since the quark-number chemical potential considered in this paper is 
flavor diagonal, we can concentrate our discussion 
on flavor-diagonal condensates.  
Under the transformation (\ref{UAtrans}), 
the flavor-diagonal quark-antiquark condensates, 
$\sigma_f = \bar{q}_f q_f$ and $\eta_f = \bar{q}_f i\gamma_5q_f$, 
are transformed into $\sigma_f^\prime = \bar{q}_f^\prime q_f^\prime$ 
and $\eta_f^\prime = \bar{q}_f^\prime i\gamma_5q_f^\prime$ as 
\begin{eqnarray}
\sigma_l
&=&
\cos{\left(\tfrac{\theta}{2}\right)}\sigma_l^\prime
+\sin{\left(\tfrac{\theta}{2}\right)}\eta_l^\prime ,
\\
\eta_l
&=&
-\sin{\left(\tfrac{\theta}{2}\right)}\sigma_l^\prime
+\cos{\left(\tfrac{\theta}{2}\right)}\eta_l^\prime ,
\\
\sigma_s
&=&
\sigma_s^\prime ,
\\
\eta_s
&=&
\eta_s^\prime
\label{Econdensates}
\end{eqnarray}
for $l=u,d$. The Lagrangian is then rewritten into 
\begin{align}
 {\cal L}  
=& {\bar q}^\prime(\gamma_{\nu} D_{\nu} + {\hat m_{0+}}+i{\hat m_{0-}}\gamma_5 
-\gamma_4 {\hat \mu} )q^\prime 
\nonumber\\
 &- G_{\rm S} \sum_{a=0}^{8} 
    [({\bar q}^\prime \lambda_a q^\prime )^2 +({\bar q }^\prime i\gamma_5 \lambda_a q^\prime  )^2] 
\nonumber\\
 &+ G_{\rm D} \Bigl[\det_{ij} {\bar q}_i^\prime (1-\gamma_5) q_j^\prime  
           +\det_{ij} {\bar q}_i^\prime (1+\gamma_5) q_j^\prime \Bigr]
\nonumber\\
&+{\cal U}(\Phi [A],\Phi^* [A],T) 
\label{L_vacuum_prime}
\end{align} 
with
\begin{eqnarray}
\hat{m}_{0+}
&=&
{\rm diag}(m_{u+},m_{d+},m_{s+})\nonumber\\
&=&
{\rm diag}\left(
\cos{\left(\tfrac{\theta}{2}\right)}m_{u},~
\cos{\left(\tfrac{\theta}{2}\right)}m_{d},~
m_{s}\right),\\
\hat{m}_{0-}
&=&
{\rm diag}(m_{u-},m_{d-},m_{s-})\nonumber\\
&=&
{\rm diag}\left(
\sin{\left(\tfrac{\theta}{2}\right)}m_{u},~
\sin{\left(\tfrac{\theta}{2}\right)}m_{d},~
0\right) .
\end{eqnarray} 

Making the mean-field approximation, one can obtain 
the mean-field Lagrangian as
\begin{eqnarray}
{\cal L}_{\rm MF}
&=&{\bar q}'(\gamma_\nu D_\nu + M_f^\prime + i\gamma_5N_f^\prime 
- \gamma_4 {\hat \mu} )q'
\nonumber\\
&&+U_M +{\cal U}(\Phi [A],\Phi^* [A],T) , 
\end{eqnarray} 
where
\bea
M_{f}^\prime &=&m_{f+}-4G_{\rm S}\sigma_{f}^\prime +  2G_{\rm D}(\sigma^{\prime}_{f^\prime }\sigma^{\prime}_{f^{\prime\prime}}-\eta^{\prime}_{f^\prime } \eta^{\prime}_{f^{\prime\prime}}), \\
N_{f}^\prime &=&m_{f-}-4G_{\rm S}\eta_{f}^\prime -  2G_{\rm D}(\sigma^{\prime}_{f^\prime } \eta^{\prime}_{f^{\prime\prime}}+\eta^{\prime}_{f^\prime } \sigma^{\prime}_{f^{\prime\prime}}) 
\eea
for $f\neq f^{\prime},~f\neq f^{\prime\prime},~f^{\prime}\neq f^{\prime\prime}$ and 
\begin{eqnarray}
 U_M 
&=&
2G_{\rm S}\sum_{f=u,d,s}({\sigma_{f}^\prime}^2+{\eta_{f}^\prime}^2)
-4G_{\rm D}\sigma_{u}^\prime \sigma_{d}^\prime \sigma_{s}^\prime
\nonumber\\
&&+4G_{\rm D}(\sigma_{u}^\prime\eta_{d}^\prime\eta_{s}^\prime
+\eta_{u}^\prime\sigma_{d}^\prime\eta_{s}^\prime
+\eta_{u}^\prime\eta_{d}^\prime\sigma_{s}^\prime).
\label{Upotential_theta}
\end{eqnarray}
Performing the path integral over the quark field, 
one can obtain the thermodynamic potential $\Omega$ (per volume) 
for finite $T$ and $\mu$: 
\begin{align}
\Omega
&= -2 \sum_{f=u,d,s} \int \frac{d^3 {\bf p}}{(2\pi)^3}
  \Bigl[ N_\mathrm{c} E_f \nonumber\\
& + \frac{1}{\beta}\ln~ [1 + 3\Phi e^{-\beta (E_f-\mu)}
    +3\Phi^* e^{-2\beta (E_f-\mu)}+ e^{-3\beta (E_f-\mu)}]
\notag\\
& + \frac{1}{\beta}\ln~ [1 + 3\Phi^*e^{-\beta (E_f+\mu)} 
    +3\Phi e^{-2\beta (E_f+\mu)}+ e^{-3\beta (E_f+\mu)}]
\Bigl]
\nonumber\\
&+ U_M +{\cal U}(\Phi,\Phi^*,T) 
\label{PNJL-Omega_theta}
\end{align}
with $E_f = \sqrt{{\bf p}^2+{M_{f}^\prime }^2+{N_{f}^\prime }^2}$.

The three-dimensional cutoff for the momentum integration 
is introduced~\cite{Matsumoto}, since this model is nonrenormalizable. 
For simplicity we assume the isospin symmetry for the ${u}$-${d}$ sector: 
$m_{l} \equiv m_{u}=m_{d}$. This three-flavor PNJL model 
has five parameters $G_{\rm S}$, $G_{\rm D}$, $m_l$, $m_s$ and $\Lambda$. 
One of the typical parameter sets is shown in Table \ref{Table_NJL}. 
These parameters are fitted to empirical values of pion decay constant and $\pi$, $K$, $\eta'$ meson masses at vacuum. 
\begin{table}[h]
\begin{center}
\begin{tabular}{ccccc}
\hline \hline
~~$m_l(\rm MeV)$~~ & ~~$m_s(\rm MeV)$~~ & ~~$\Lambda(\rm MeV)$~~ & ~~$G_{\rm S} \Lambda^2$~~ & ~~$G_{\rm D} \Lambda^5$~~
\\
\hline
$5.5$ & $140.7$ & $602.3$ & $1.835$ & $12.36$
\\
\hline \hline
\end{tabular}
\caption{
Summary of the parameter set in the NJL sector taken 
from Ref.~\cite{Rehnberg}.
\label{Table_NJL}
}
\end{center}
\end{table}

For imaginary $\mu$, $\Omega$ is invariant 
under the extended ${\mathbb Z}_3$ transformation~\cite{Sakai}, 
\begin{align}
&e^{\pm \mu/T} \to e^{\pm \mu/T} e^{\pm i{2\pi k\over{3}}},\quad  
\Phi  \to \Phi e^{-i{2\pi k\over{3}}}, 
\notag\\
&\Phi ^{*} \to \Phi ^{*} e^{i{2\pi k\over{3}}} ,
\label{eq:K2}
\end{align}
with integer $k$. 
This invariance ensures the Roberge-Weiss periodicity~\cite{RW} 
in the imaginary chemical potential region~\cite{Sakai}. 
Any reliable model should have this extended ${\mathbb Z}_3$ symmetry, 
when imaginary $\mu$ is taken in the model. This is a good test for checking 
the reliability of the model. The PNJL model has 
the extended ${\mathbb Z}_3$ symmetry~\cite{Sakai,Matsumoto}. 

The four-quark vertex $G_{\rm S}$ is originated in a one-gluon exchange 
between quarks and its higher-order diagrams. 
If the gluon field $A_{\nu}$ has a vacuum expectation value 
$\langle A_{0} \rangle$ in its time component, 
$A_{\nu}$ is coupled to $\langle A_{0} \rangle$ and then to $\Phi$ 
through $L$.
Hence we can modify $G_{\rm S}$ into an effective vertex 
$G_{\rm S}(\Phi)$ depending on $\Phi$~\cite{Kondo}. 
The effective vertex $G_{\rm S}(\Phi)$ is called the entanglement vertex and 
the model with this vertex is the EPNJL model. 
It is expected that $\Phi$ dependence of $G_{\rm S}(\Phi )$ 
will be determined in future by the accurate method 
such as the exact renormalization group method~\cite{Braun,Kondo,Wetterich}. 
In this paper, however, we simply assume the following form for 
$G_{\rm S}(\Phi )$: 
\begin{eqnarray}
G_{\rm S}(\Phi)=G_{\rm S}[1-\alpha_1\Phi\Phi^*-\alpha_2(\Phi^3+\Phi^{*3})]. 
\label{entanglement-vertex}
\end{eqnarray}
This form preserves the chiral symmetry, the charge conjugation ($C$) 
symmetry~\cite{Kouno} and the extended $\mathbb{Z}_3$ symmetry~\cite{Sakai}. 
This entanglement vertex modifies the mesonic potential $U_M$, 
the dynamical quark masses $M_{f}'$ and $N_{f}'$. 
This is the three-flavor version of the EPNJL model. 
In principle, $G_{\rm D}$ can depend on $\Phi$, but 
$\Phi$ dependence of $G_{\rm D}$ is found to yield qualitatively 
the same effect on the phase diagram as that of $G_{\rm S}$. 
Following Ref.~\cite{Sasaki-T_Nf3}, we neglect 
$\Phi$ dependence of $G_{\rm D}$ as a simple setup. 
In the present analysis, $\Phi$ dependence of $G_{\rm D}$ is thus 
renormalized in $\alpha_1$ and $\alpha_2$. 
The EPNJL model thus constructed keeps the extended ${\mathbb Z}_3$ 
symmetry. 

The parameters $\alpha_1$ and $\alpha_2$ in \eqref{entanglement-vertex} 
are fitted to two results of LQCD at finite $T$; 
one is the result of 2+1 flavor LQCD at $\mu=0$~\cite{YAoki} 
that the chiral transition is crossover at the physical point 
and another is the result of degenerate three-flavor LQCD 
at $\mu=i T \pi$~\cite{FP2010} that the order of the RW phase transition 
at the RW endpoint is first order for small and large quark masses 
and second order for intermediate quark masses. 
The parameter set $(\alpha_1, \alpha_2)$ thus determined is located in 
the triangle region 
\bea
\{-1.5\alpha_1+0.3 < \alpha_2 <-0.86\alpha_1+0.32,~\alpha_2 >0\}. 
\label{triangle}
\eea
In this paper we take $\alpha_1=0.25,~\alpha_2=0.1$ 
as a typical example, following Ref.~\cite{Sasaki-T_Nf3}. 

The classical variables $X=\Phi$, ${\Phi}^*$, $\sigma_f$ and $\eta_f$ 
are determined by the stationary conditions 
\begin{align}
\partial \Omega/\partial X=0. 
\label{eq:SC}
\end{align}
The solutions to the stationary conditions do not give the global minimum of $\Omega$ necessarily. 
There is a possibility that they yield a local minimum or even a maximum. We then have checked that the solutions yield the global minimum when  the solutions $X(T, \mu, \theta )$ are inserted into (\ref{PNJL-Omega_theta}).  

\section{Numerical results}
\label{Numerical results}

In this section we show numerical results for the original condensates $(\sigma_{f},\eta_{f},\Phi)$, 
since this makes our discussion transparent. 
Under the parity transformation, $\sigma_f$, $\eta_f$ and $\Phi$ are 
transformed into $\sigma_f$, $-\eta_f$ and $\Phi$, respectively. 
This means that $\eta_f$ is $\theta$ odd while 
$\sigma_f$ and $\Phi$ are $\theta$ even, since 
the Lagrangian is invariant under the combination of 
the parity transformation and the transformation $\theta \to -\theta$. 
Thus $\eta_f$ is an order parameter of the spontaneous parity breaking, 
while  $\sigma_f$ and $\Phi$ are approximate order parameters 
of the chiral and the deconfinement transitions, respectively. 
As an approximate order parameter of the chiral transition, 
$\sigma_l \equiv \sigma_u=\sigma_d$ is more proper than $\sigma_s$, 
since $m_l \ll m_s$.

\subsection{Thermodynamics at $\mu =0$}
In this subsection, we consider the case of $\mu =0$ where 
charge conjugation symmetry ($C$) is exact. Meanwhile, 
parity symmetry ($P$) is exact only at $\theta=0,~\pm \pi$, 
since $e^{i\theta}$ agrees with $e^{-i\theta}$ 
in \eqref{L_vacuum} when $\theta=0,~\pm \pi$. 

\begin{figure}[t]
\begin{center}
\hspace{-10pt}
 \includegraphics[width=0.30\textwidth,bb= 90 50 255 220,clip]{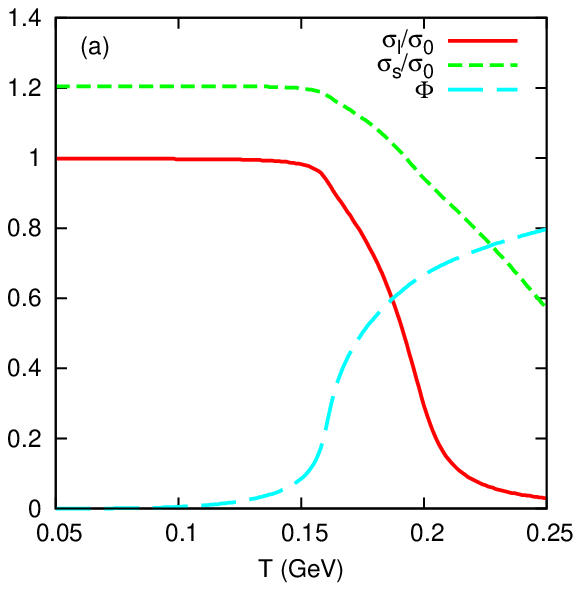}
 \includegraphics[width=0.30\textwidth,bb= 90 50 255 220,clip]{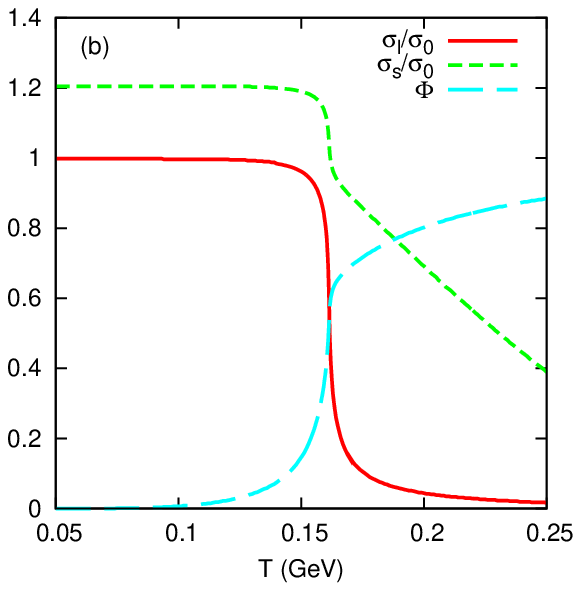}
\end{center}
\vspace{-10pt}
\caption{
$T$ dependence of the order parameters at $\theta=0$ and $\mu =0$ in 
(a) the PNJL model and (b) the EPNJL model. 
The solid, dashed and dotted lines represent 
$\sigma_l$, $\sigma_s$ and $\Phi$, respectively, 
where $\sigma_l$ and $\sigma_s$ are normalized by 
$\sigma_0=\sigma_{l}(T=\mu=\theta=0)$. 
}
\label{order_theta000_mu000}
\end{figure}

Figure \ref{order_theta000_mu000} shows $T$ dependence of $\sigma_l$, $\sigma_s$ and $\Phi$ at $\theta=\mu =0$, 
where $\sigma_l$ and $\sigma_s$ are normalized by $\sigma_0=\sigma_{l}$ at $T=\mu=\theta=0$. 
Here $\sigma_l$ and $\Phi$ describe the chiral and deconfinement transitions, 
respectively. In the PNJL model of panel (a), 
the chiral restoration transition takes place after the deconfinement transition. 
In the EPNJL model of panel (b), meanwhile, both the transitions occur simultaneously. 
In the EPNJL model, $\sigma_s$ decreases rapidly near the pseudocritical temperature $T_{\rm c}=160$~MeV, 
but goes down gradually above $T_{\rm c}$. 
The rapid change of $\sigma_s$ comes from that of $\Phi$. 
For both the PNJL and EPNJL models, $\eta_l$ and $\eta_s$ are zero 
at any $T$. The $P$ symmetry is thus always preserved when $\theta=0$. 

\begin{figure}[t]
\begin{center}
\hspace{-10pt}
 \includegraphics[width=0.3\textwidth,bb= 70 50 255 220,clip]{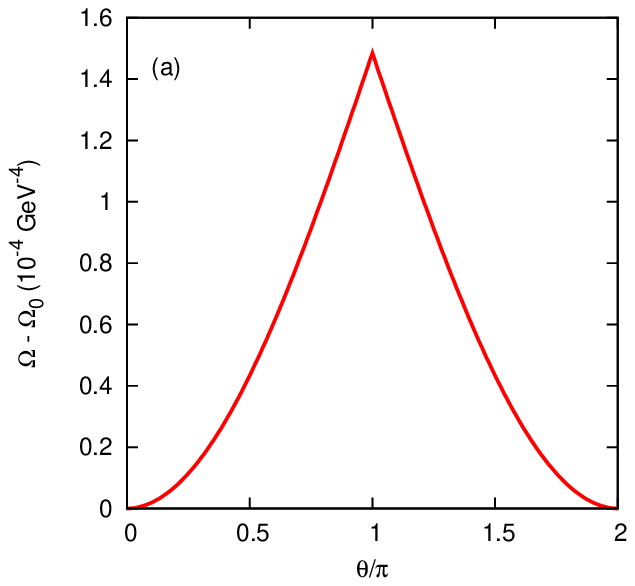} 
 \includegraphics[width=0.3\textwidth,bb= 70 50 255 220,clip]{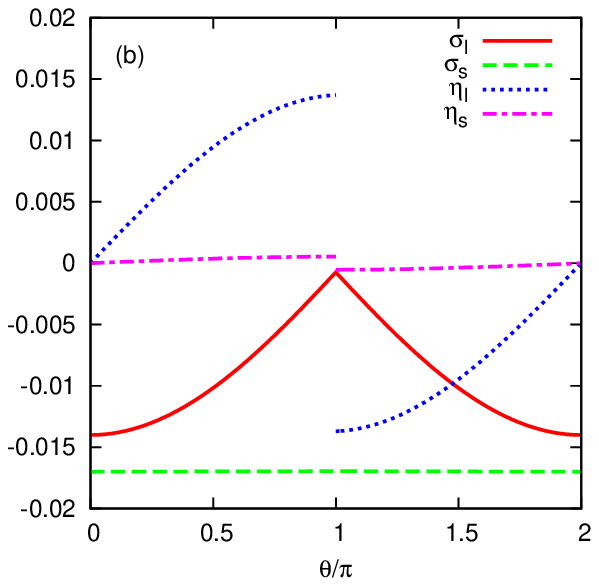} 
\vspace{-15pt}
\end{center}
\caption{$\theta$ dependence of (a) $\Omega$ and (b) the order parameters 
at $T=\mu =0$ in the EPNJL model. 
In panel (a), $\Omega_0\equiv \Omega(\theta =0)$ is subtracted from $\Omega$. 
See the legend for the definition of lines.
}
\label{T000_mu000}
\end{figure}

Figure \ref{T000_mu000} shows $\theta$ dependence of $\Omega$ 
and the order parameters at $T=\mu =0$ in the EPNJL model; 
note that the EPNJL model agrees 
with the PNJL model at $T=0$, since $G_{\rm S}(\Phi) =G_{\rm S}$ there 
because of $\Phi =0$. 
As shown in panel (a), $\Omega$ is $\theta$ even and has a cusp 
at $\theta=\pi$. This indicates that a first-order phase transition 
takes place at $T=\mu=0$ and $\theta=\pi$. 
As shown in panel (b), meanwhile, the $\eta_f$ are $\theta$ odd, 
while $\sigma_f$ and $\Phi$ are $\theta$ even. The condensate 
$\eta_l$ and $\eta_s$ have jumps at $\theta=\pi$, indicating that 
the first-order transition mentioned above is 
the spontaneous parity breaking.  
This is nothing but the Dashen phenomena~\cite{Dashen}.   

\begin{figure}[t]
\begin{center}
\hspace{-10pt}
 \includegraphics[width=0.3\textwidth,bb= 85 50 255 220,clip]{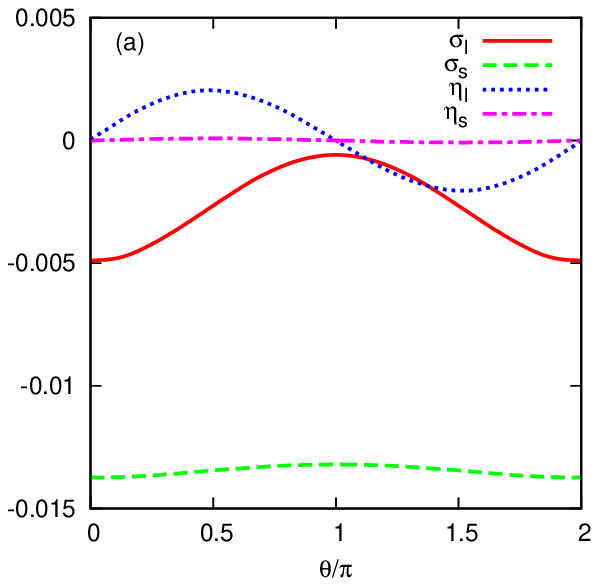} 
 \includegraphics[width=0.3\textwidth,bb= 85 50 255 220,clip]{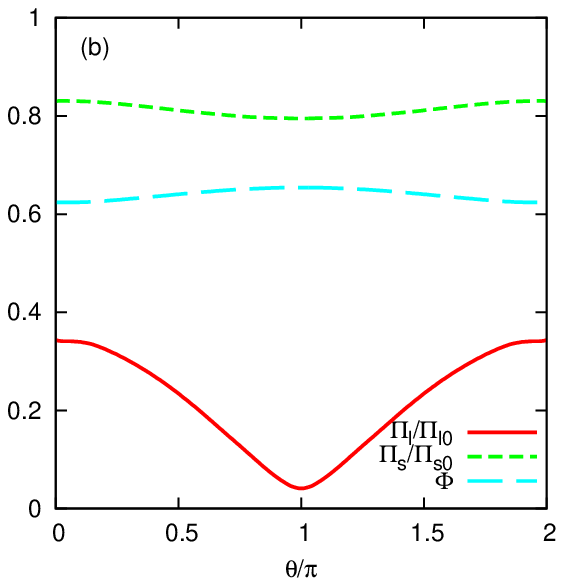} 
\vspace{-15pt}
\end{center}
\caption{
$\theta$ dependence of (a) the order parameters and (b) 
the effective quark mass $\Pi_f$ at $T=163$~MeV and $\mu =0$ 
in the EPNJL model. 
In panel (b), $\Pi_f$ is normalized by the value at $T=\mu=\theta =0$ and 
the normalized $\Pi_f$ is compared with the Polyakov loop $\Phi$. 
See the legend for the definition of lines.
}
\label{Order_t163_mu0_thetadep}
\end{figure}

Figure \ref{Order_t163_mu0_thetadep} shows $\theta$ dependence of the order parameters and the effective quark mass 
$\Pi_f\equiv \sqrt{{M_f^\prime }^2+{N_f^\prime }^2}$ at $T=163$~MeV and $\mu =0$ in the EPNJL model. 
For this higher temperature, 
the Dashen phenomena do not take place at $\theta= \pi$. 
Actually $\eta_l$ and $\eta_s$ vanish there, although they become finite at $\theta \neq 0,~\pi,~2\pi$ where $P$ is not an exact symmetry. 
The other order parameters, $\sigma_f$ and $\Phi$, are smooth periodic functions of $\theta$. 
The Polyakov loop $\Phi$ becomes maximum at $\theta =\pi$, 
since the effective quark mass $\Pi_f$ becomes minimum and 
the thermal factor $\exp(-\beta E_f)$ is maximized in \eqref{PNJL-Omega_theta}. 

\begin{figure}[t]
\begin{center}
\hspace{-10pt}
 \includegraphics[width=0.3\textwidth,bb= 90 50 255 220,clip]{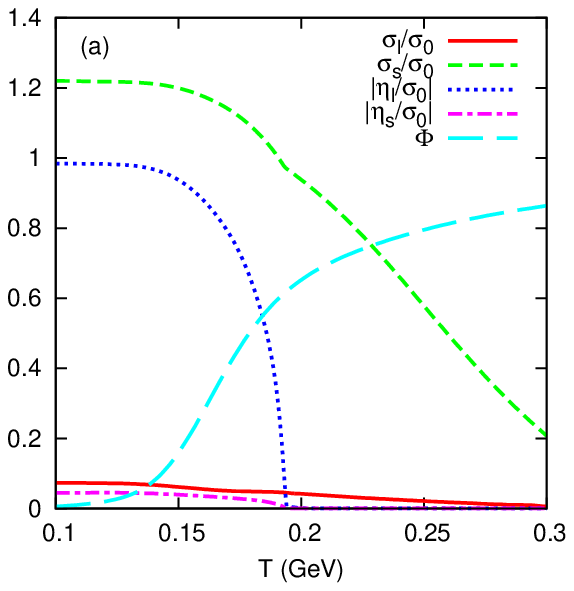} 
 \includegraphics[width=0.3\textwidth,bb= 90 50 255 220,clip]{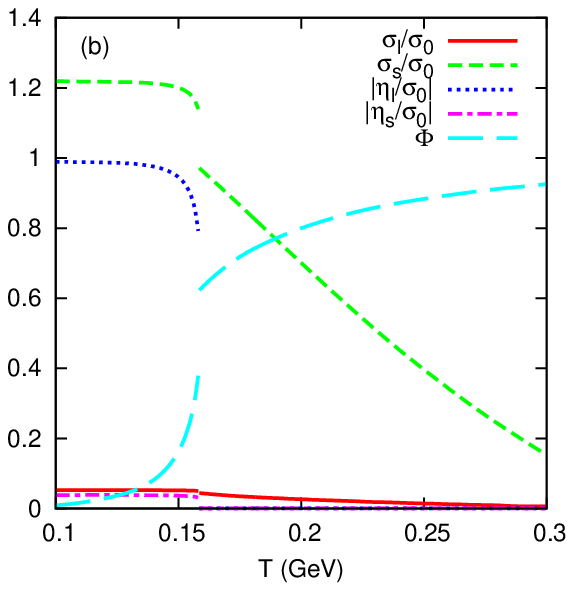} 
\vspace{-15pt}
\end{center}
\caption{
$T$ dependence of the order parameters at $\theta =\pi$ and $\mu =0$ in 
(a) the PNJL model and (b) the EPNJL model. 
See the legend for the definition of lines. 
}
\label{Order_thetapi_mu000_Tdep}
\end{figure}

Figure \ref{Order_thetapi_mu000_Tdep} shows $T$ dependence of the order parameters at $\theta =\pi$ and $\mu=0$. 
Comparing this figure with Fig.~\ref{order_theta000_mu000}, one can also see $\theta$ dependence of the order parameters. 
In the PNJL model of panel (a), $|\eta_l|$ and $|\eta_s|$ are finite below the critical temperature $T_P=194$~MeV and vanish above $T_P$. 
Thus the $P$ symmetry is broken at smaller $T$, but restored at higher $T$. 
In the two-flavor PNJL model, 
this $P$ restoration is second order~\cite{Sakai_CP}. 
This is the case also for the present 2+1 flavor PNJL model. 
The second order $P$ restoration induces cusps in $|\sigma_l|$ and 
$|\sigma_s|$ when $T=T_P$, although the cusp is weak in $|\sigma_s|$. 
This propagation of the cusp can be understood 
by the extended discontinuity theorem of Ref.~\cite{Kashiwa5}. 
In the EPNJL model of panel (b), the $P$ restoration occurs at $T_P=158$~MeV 
as the first-order transition. The same property is seen in 
the two-flavor EPNJL model~\cite{Sakai_CP}. 
The first-order $P$ restoration generates gaps in $|\sigma_l|$ and 
$|\sigma_s|$ when $T=T_P$, although the gap is tiny in $|\sigma_s|$. 
This propagation of the gap can be understood by the discontinuity 
theorem by Barducci, Casalbuoni, Pettini and Gatto~\cite{BCPG}. 
Thus the Dashen phenomena are seen only at lower $T$, and the order of 
the $P$ violation at the critical temperature $T_P$ depends on 
the effective model taken. 

Theoretical prediction on the critical temperature of 
the chiral transition at $\theta=0$ and $\mu=0$
and the P restoration at $\theta =\pi$ and $\mu=0$ 
is tabulated in Table \ref{table-tc}. 
At $\theta =0$, the chiral transition is crossover in all of the NJL, PNJL, and EPNJL models
At $\theta=\pi$, the order of the P restoration is first order 
in the EPNJL model, but it is second order in the PNJL and NJL models. 
\begin{table}[t]
\begin{center}
\begin{tabular}{cll}
\hline \hline
Model& ~~$\theta =0$~~ & ~~$\theta =\pi$~~\\
\hline
NJL&~~$177$~(crossover)~~ & ~~$170$~(2nd order)~~ \\
\hline
PNJL&~~$200$~(crossover)~~ & ~~$194$~(2nd order)~~ \\
\hline
EPNJL&~~$162$~(crossover)~~ & ~~$158$~(1st order)~~ \\
\hline \hline
\end{tabular}
\caption{
Theoretical prediction on the critical temperature of 
the chiral transition at $\theta=0$ and $\mu=0$
and the P restoration at $\theta =\pi$ and $\mu=0$. 
The values are shown in units of MeV. 
}
\label{table-tc}
\end{center}
\end{table}

\subsection{Thermodynamics at $\mu > 0$}
In this subsection, we consider the case of $\mu > 0$ where 
$C$ symmetry is not exact. 
In general, the relation $\Phi =\Phi^*$ is not satisfied for finite $\mu$, 
although $\Phi$ and $\Phi^*$ are real~\cite{Sakai_phase}. 
This situation makes numerical calculations quite time-consuming. 
However, the deviation $\Phi-\Phi^*$ is known 
to be very small~\cite{Sakai_phase}. For this reason, 
the assumption $\Phi =\Phi^*$ has been used in many calculations. 
Therefore we use the assumption also in this paper. 

\begin{figure}[t]
\begin{center}
\hspace{-10pt}
 \includegraphics[width=0.3\textwidth,bb= 90 50 255 220,clip]{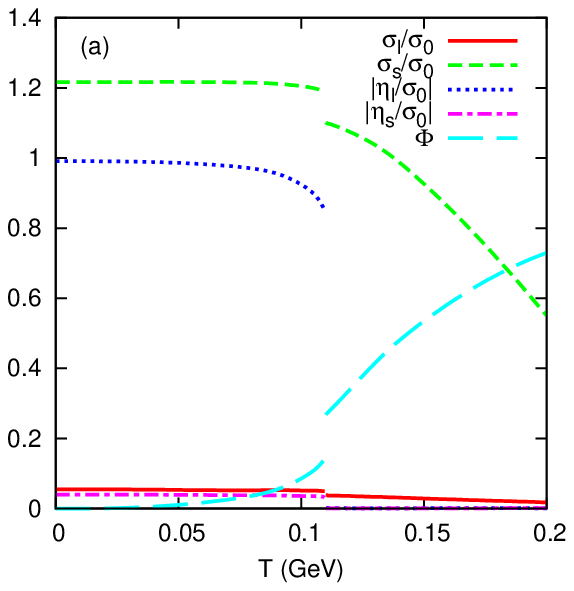} 
 \includegraphics[width=0.3\textwidth,bb= 90 50 255 220,clip]{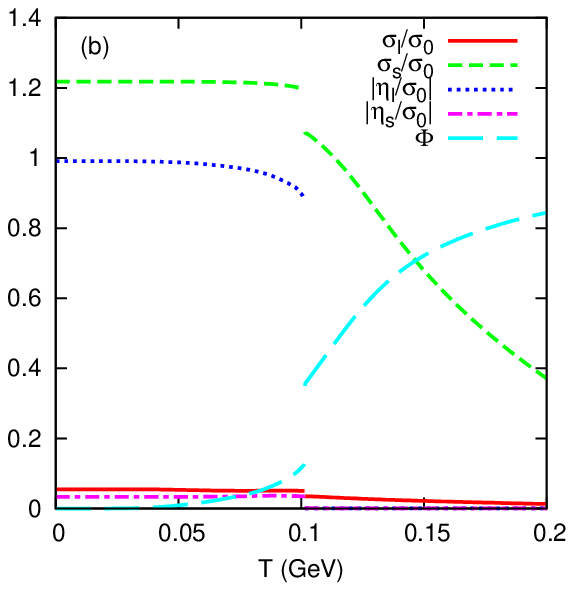} 
\vspace{-15pt}
\end{center}
\caption{
$T$ dependence of the order parameters at $\theta =\pi$ and $\mu =300$~MeV 
in (a) the PNJL model and (b) the EPNJL model. 
See the legend for the definition of lines.
}
\label{Order_thetapi_mu300_Tdep}
\end{figure}

Figure \ref{Order_thetapi_mu300_Tdep} represents $T$ dependence 
of the order parameters at $\theta =\pi$ and $\mu =300$~MeV 
in the PNJL and EPNJL models. 
The $P$ restoration takes place at high $T$, since $\eta_l$ and $\eta_s$ are zero there.
The critical temperature of the $P$ restoration is $T_P=110$~MeV for 
the PNJL model and $T_P=99$~MeV for the EPNJL model. 
For $\mu=300$~MeV, the order of the $P$ restoration at $T=T_P$ is first order 
in both the PNJL and EPNJL models. 
Thus the quark-number chemical potential $\mu$ lowers $T_P$ and 
makes the $P$ restoration sharper. 

\begin{figure}[t]
\begin{center}
\hspace{-10pt}
 \includegraphics[width=0.3\textwidth,bb= 50 60 210 170,clip]{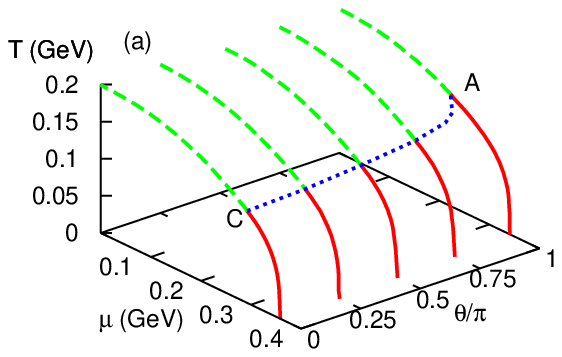} 
 \includegraphics[width=0.3\textwidth,bb= 50 60 210 170,clip]{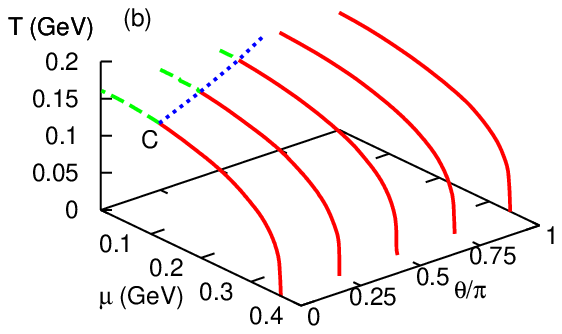} 
\vspace{-15pt}
\end{center}
\caption{
Phase diagram of the chiral transition in the $\mu$-$\theta$-$T$ space. 
Panel (a) shows a result of the PNJL model and panel (b) corresponds to a 
result of the EPNJL model. 
}
\label{3dim_plot}
\end{figure}

Figure \ref{3dim_plot} shows the phase diagram of the chiral transition 
in the $\mu$-$\theta$-$T$ space. 
The diagram is mirror symmetric with respect 
to the $\mu$-$T$ plane at $\theta =0$, so the diagram is plotted only 
at $\theta \ge 0$.   
Panels (a) and (b) correspond to results of the PNJL and EPNJL models, 
respectively.
In the $\mu$-$T$ plane at $0 \le \theta <\pi$, 
the solid line stands for the first-order chiral transition, 
while  the dashed line represents the chiral crossover. 
The meeting point between the solid and dashed lines is a critical endpoint 
(CEP) of second order. 
Point C is a CEP in the $\mu$-$T$ plane at $\theta=0$~\cite{AY,Barducci_CEP}. 
For both the PNJL and EPNJL models, the location of CEP in the $\mu$-$T$ 
plane moves to higher $T$ and lower $\mu$ as $\theta$ increases from 0 
to $\pi$. 

In the $\mu$-$T$ plane at $\theta=\pi$, $P$ symmetry is exact and 
hence we can consider the spontaneous breaking of $P$ symmetry in addition to 
the chiral transition. 
For the PNJL model of panel (a), 
both the first-order chiral transition and the first-order $P$ restoration 
take place simultaneously, and the second-order $P$ 
restoration and the chiral crossover coincide with each other. 
The first-order and the second-order $P$ transition line are depicted 
by the solid and dashed lines, respectively. 
The meeting point A is a tricritical point (TCP) of the $P$-restoration 
transition. 
For the EPNJL model of panel (b), the chiral and the $P$ restoration 
transition are always first order and hence there is no TCP. 

In the PNJL model of panel (a), the dotted line from point C to point A 
is a trajectory of CEP as $\theta$ increases from 0 to $\pi$. 
Thus the second-order chiral transition line ends up with point A. 
This means that the CEP (point C) at $\theta=0$ is a remnant of 
the TCP (point A) of $P$ restoration at $\theta=\pi$. 
In the EPNJL model of panel (b), 
no TCP and then no CEP appears in the $\mu$-$T$ plane at $\theta=\pi$. 
The second-order chiral-transition line (dashed line) starting 
from point C never reaches the $\mu$-$T$ plane at $\theta=\pi$. 

\begin{figure}[t]
\begin{center}
\hspace{-10pt}
 \includegraphics[width=0.45\textwidth,bb= 70 50 235 170,clip]{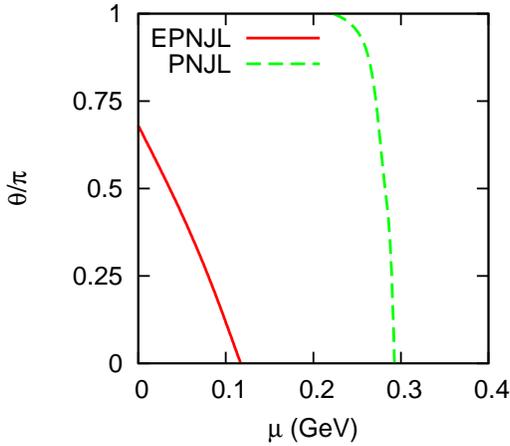} 
\vspace{-15pt}
\end{center}
\caption{
The projection of the second-order chiral-transition line 
in the $\mu$-$\theta$-$T$ space on the $\mu$-$\theta$ plane.
See the legend for the definition of line. 
}
\label{CEP_line}
\end{figure}

Figure \ref{CEP_line} snows the projection of 
the second-order chiral-transition line in the $\mu$-$\theta$-$T$ space 
on the $\mu$-$\theta$ plane. 
The solid (dashed) line stands for the projected line 
in the EPNJL (PNJL) model. 
The first-order transition region exists on the right-hand side of 
the line, while the left-hand side corresponds to the chiral crossover 
region. The first-order transition region is much wider in the EPNJL model 
than in the PNJL model. In the EPNJL model, eventually, 
the chiral transition becomes first order even at $\mu=0$ when $\theta$ is 
large. 

\subsection{The sign problem on LQCD with finite $\theta$}
\begin{figure}[t]
\begin{center}
\hspace{-10pt}
 \includegraphics[width=0.3\textwidth,bb= 85 50 255 220,clip]{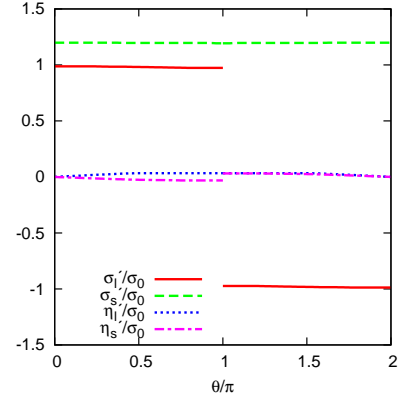} 
\vspace{-15pt}
\end{center}
\caption{
$\theta$ dependence of the order parameters, 
$\sigma_l^\prime$, $\sigma_s^\prime$, $\eta_l^\prime$ 
and $\eta_s^\prime$, at $T=\mu =0$ calculated with the EPNJL model.
See the legend for the definition of lines.
}
\label{order-prime}
\end{figure}

In the PNJL Lagrangian \eqref{L_vacuum_prime} after the 
transformation (\ref{UAtrans}), 
$\theta$ dependence appears only at the light quark mass terms, 
$m_l \cos(\theta/2)$ and $m_l \sin(\theta/2)$. 
These terms are much smaller than $\Lambda_{QCD}$ as a typical scale of QCD. 
This means that the condensates, $\sigma_l^\prime$, $\sigma_s^\prime$, 
$\eta_l^\prime$  and $\eta_s^\prime$, have weak $\theta$ dependence. 
This statement is supported by the results of the PNJL calculations   
shown in Fig.~\ref{order-prime}. 

The sign problem is induced by the $\theta$ odd  $m_l \sin(\theta/2)$ term. 
The $\theta$-odd ($P$-odd) condensates, $\eta_l^\prime$ and $\eta_s^\prime$, 
are generated by the $\theta$-odd mass term. 
One can see in Fig.~\ref{order-prime} 
that the  $\theta$-odd condensates are much smaller than the 
$\theta$-even condensates, $\sigma_l^\prime$ and $\sigma_s^\prime$. 
This fact indicates that effects of the $\theta$-odd mass term are negligible. 
Actually, if the term is neglected, the $\theta$-even condensates  
change only within the thickness of line, while 
the $\theta$-odd condensates vanish. The neglect of the $\theta$-odd mass 
is thus a good approximation. 

The validity of the approximation can be shown more explicitly 
in the following way. 
The $\theta$-odd ($P$-odd) condensates, $\eta_l^\prime$ and $\eta_s^\prime$, 
are zero at $\theta=0$, since the $\theta$-odd mass vanishes there. 
The weak $\theta$ dependence of $\eta_l^\prime$ and $\eta_s^\prime$ 
guarantees that $\eta_l^\prime$ and $\eta_s^\prime$ are small for any 
$\theta$. Setting $\eta_l^\prime=\eta_s^\prime=0$ in $M'_l$ and $N_l'$ leads to
\begin{eqnarray}
M_l^\prime
&=&
\cos{\left(\tfrac{\theta}{2}\right)}m_l
-4G_{\rm S}\sigma^{\prime}_l+2G_{\rm D}\sigma^{\prime}_s\sigma^{\prime}_l,
\label{Mlp2}
\\
N_l^\prime
&=&
\sin{\left(\tfrac{\theta}{2}\right)}m_l , 
\label{Nlp2}
\end{eqnarray}
where $M_l^\prime \approx \Lambda_{\rm QCD}$ and $N_l^\prime \approx m_l$.
Since the thermodynamic potential is a function of $M_l^{\prime 2}+N_l^{\prime 2}$, the term $N_l^{\prime 2}$ is negligible compared 
with $M_l^{\prime 2}$.

In LQCD, the vacuum expectation value of operator ${\cal O}$ is obtained by 
\begin{eqnarray}
\langle{\cal O}\rangle
&=&
\int\mathcal{D}A
{\cal O}\left( {\rm det}{\cal M}_l(\theta )\right) ^2{\rm det}{\cal M}_s e^{-S_g}
\\
&=&
\int\mathcal{D}A
{\cal O}'\left( {\rm det}{\cal M}_l'(\theta ) \right) ^2{\rm det}{\cal M}_s e^{-S_g}
\label{reweighting-method}
\end{eqnarray}
with the gluon part $S_g$ of the QCD action and  
\begin{equation}
{\cal O}'\equiv
{\cal O}\frac{\left( {\rm det}{\cal M}_l(\theta )\right) ^2}{\left( {\rm det}{\cal M}_l'(\theta )\right) ^2} ,
\end{equation}
where ${\rm det}{\cal M}'_l(\theta )$~ is the Fermion determinant in which 
the $\theta$-odd mass is neglected and hence has no sign problem. 
As mentioned above, one can assume that 
\begin{equation}
\frac{{\rm det}{\cal M}_l(\theta )}{{\rm det}{\cal M}'_l(\theta )}\approx 1 .
\end{equation}
Thus the reweighting method defined by \eqref{reweighting-method} 
may work well. 
In the $\theta$-even mass, $m_l \cos(\theta/2)$, 
the limit of $\theta=\pi$ corresponds to the limit of $m_l=0$ 
with $m_s$ fixed. 
Although the limit is hard to reach, one can analyze the dynamics at least 
at small and intermediate $\theta$.

\section{Summary}
\label{Summary}
We have investigated effects of the theta vacuum on the 
QCD phase diagram for the realistic 2+1 flavor system, 
using the three-flavor PNJL and EPNJL models. 
The effects can be easily understood by 
the $SU_{\rm A}(3) \otimes U_{\rm A}(1)$ transformation \eqref{UAtrans}. 
After the transformation, the $\theta$-odd mass, $m_l \sin(\theta/2)$, 
little affects the dynamics, so that the dynamics is mainly governed 
by the $\theta$-even mass, $m_l \cos(\theta/2)$. 
In the $\theta$-even mass, 
the increase of $\theta$ corresponds to the decrease of $m_l$ with 
$m_s$ fixed. 
This means that the chiral transition becomes strong 
as $\theta$ increases. This is true in the results of both PNJL and EPNJL 
calculations. Particularly in the EPNJL model that is more reliable 
than the PNJL model, the transition becomes first-order even at $\mu=0$ 
when $\theta$ is large. This result is important. 
If the chiral transition becomes first order at $\mu=0$, 
it will change the scenario of cosmological evolution. 
For example, the first-order transition allows us 
to think the inhomogeneous Big-Bang nucleosynthesis model or 
a new scenario of baryogenesis. 

Using the fact that the $\theta$-odd mass is negligible, we have proposed 
a way of circumventing the sign problem on LQCD with finite $\theta$. 
The reweighting method by defined \eqref{reweighting-method} 
may allow us to do LQCD calculations and get definite results on the dynamics 
of the $\theta$ vacuum. 

\begin{acknowledgments}
The authors thank A. Nakamura, T. Saito, K. Nagata and K. Nagano for useful discussions. 
H.K. also thanks M. Imachi, H. Yoneyama, H. Aoki and M. Tachibana for useful discussions. 
T.S and Y.S. are supported by JSPS.
The numerical calculations were performed on the HITACHI SR16000 at Kyushu University and the NEC SX-9  at CMC, Osaka University.
\end{acknowledgments}


\end{document}